\documentclass[12pt,a4paper]{article}
\begin{document}
\begin{center}
{\LARGE \bf Entanglement Teleportation Through 1D Heisenberg
Chain}

\vskip0.8cm {\it Xiang Hao and Shiqun Zhu*}

\vskip0.2cm
{\it China Center of Advanced Science and Technology(World Laboratory),\\
P. O. Box 8730, Beijing 100080, People's Republic of China\\
and\\
School of Physical Science and Technology, Suzhou University,\\
Suzhou, Jiangsu 215006, People's Republic of China**}
\end{center}

\begin{abstract}

\qquad Information transmission of two qubits through two
independent 1D Heisenberg chains as a quantum channel is analyzed.
It is found that the entanglement of two spin-$\frac 12$ quantum
systems is decreased during teleportation via the thermal mixed
state in 1D Heisenberg chain. The entanglement teleportation will
be realized if the minimal entanglement of the thermal mixed state
is provided in such quantum channel. High average fidelity of
teleportation with values larger than $2\!/3$ is obtained when the
temperature {\it T} is very low. The mutual information
$\mathcal{I}$ of the quantum channel declines with the increase of
the temperature and the external magnetic field. The entanglement
quality of input signal states cannot enhance mutual information
of the quantum channel.

\end{abstract}


* Corresponding author, email: szhu@suda.edu.cn

** Mailing address

\newpage

\section{Introduction}

\qquad Quantum teleportation originally proposed by Bennett {\it
et. al.} [1] has developed rapidly in recent years. The maximally
entangled state shared between Alice and Bob plays a crucial role
in the standard teleportation method. The entanglement is
essential in quantum information because of its unique advantages
that can be applied to many fields, such as quantum teleportation
[1], quantum cryptograph [2], etc. However, the decoherence from
environment always impacts on the degree of entanglement, so the
resource of maximally entangled states is hard to prepare in a
real experiment. Certainly, a mixed entangled state as the
resource is approximately near to the real circumstances. For
experimental investigations of quantum computation and quantum
communication, the thermal equilibrium state in Heisenberg model
is one fundamental sort of mixed and entangled states [3, 4].
Recently, Bowen {\it et. al.} [5] and Horodeckies [6] have
suggested that teleportation as the resource of mixed states via
the standard method, i. e., Bell-type measurements and Pauli
rotations, is equivalent to a general depolarising channel [5-7].
In some schemes of teleportation using Heisenberg model, half of
two qubits can be teleported by the sole two-qubit mixed state
[5]. In other schemes, entanglement of a special mixed state, i.
e., Werner state, can be transferred through a pair of two-qubit
mixed states [8, 9]. However, the entanglement teleportation of
the whole two-qubit system as the resource of the thermal mixed
state needs to be investigated. Meanwhile, to see more clearly the
possible applications of Heisenberg model in quantum
teleportation, mutual information of the quantum channel also
needs to be analyzed.

\qquad In this paper, the information transmission by a pair of
thermal mixed states in 1D Heisenberg $XXX$ chains is
investigated. The minimal entanglement in the quantum channel is
needed to transfer entanglement information. The high average
fidelity of teleportation is obtained when the temperature is very
low and the magnetic field is weak. The entanglement quality of
input signal states and the mutual information are analyzed. In
section 2, entanglement of the thermal mixed state in a 1D
Heisenberg $XXX$ chain is presented. The entanglement
teleportation of two-qubit pure states and fidelity are derived.
In section 3, the mutual information in such quantum channel is
analyzed. A discussion concludes the paper.

\section{Entanglement teleportation and fidelity}

\qquad In an isotropic Heisenberg model [3], there are $N$
spin-$\frac 12$ units in the quantum system with coupling constant
{\it J}. For the case of $N=2$, the Hamiltonian of the system in
the magnetic field {\it B} can be written as
\begin{equation}\label{1} H=\frac
12B(\sigma^{1}_{z}+\sigma^{2}_{z})+\frac J2(
\sigma^{1}_{x}\otimes\sigma^{2}_{x}+\sigma^{1}_{y}\otimes\sigma^{2}_{y}+
\sigma^{1}_{z}\otimes\sigma^{2}_{z}),
\end{equation}
where the periodic boundary condition $N=N+1$ is adopted, $
\sigma^{i}_{x}, \sigma^{i}_{y}, \sigma^{i}_{z}$ are Pauli rotation
operators for the $i$th qubit. The density matrix of the thermal
entangled state at the equilibrium temperature {\it T\ } can be
expressed as $\rho^{c}=\frac {1}{\textstyle z}\exp(-H/kT)$. Here,
$z$ is the partition function with $z=tr(\exp(-H/kT))$, $k$
denotes the Boltzman constant and is assumed to be one.

\qquad Assuming $\{|0\rangle, |1\rangle\}$ are the eigenstates of
$ \sigma_{z}$, the density matrix $\rho^{c}$ in the basis of
$\{|00\rangle, |01\rangle, |10\rangle, |11\rangle\}$ can be
expressed as,
\begin{equation}\label{2}
\rho^{c}= \frac 1z\left(\begin{array}{cccc}
e^{\frac {-B-\frac J2}{T}}&0&0&0\\
0&\frac 12(e^{-\frac {J}{2T}}+e^{\frac {3J}{2T}})&\frac 12(e^{-\frac {J}{2T}}-e^{\frac {3J}{2T}})&0\\
0&\frac 12(e^{-\frac {J}{2T}}-e^{\frac {3J}{2T}})&\frac 12(e^{-\frac {J}{2T}}+e^{\frac {3J}{2T}})&0\\
0&0&0&e^{\frac {B-\frac J2}{T}}
\end{array}
\right),
\end{equation}
where $z=\exp(\frac {B-\frac J2}{T})+\exp(\frac {-B-\frac
J2}{T})+\exp(-\frac {J}{2T})+\exp(\frac {3J}{2T})$. By the method
introduced in Refs. [8-11], the amount of entanglement of the
thermal state is
\begin{equation}\label{3}
E(\rho^{c})=\max\{-2\sum_{i}\lambda^{-}_{i},0\},
\end{equation}
where $\lambda^{-}_{i}$ is the $i$th negative eigenvalue of
$(\rho^{c})^{T}$ which is the partial transposition matrix of
$\rho^{c}$. In a 1D Heisenberg XXX chain, the entanglement of
$\rho^{c}$ is
\begin{equation}\label{4} E=\max\{\frac {2\exp(-\frac
{J}{2T})\cosh(\frac {B}{T})}{z}[\sqrt{1+\frac {(\exp(\frac
{2J}{T})-1)^{2}-4}{4\cosh^{2}(\frac {B}{T})}}-1],0\}.
\end{equation}
It is found that there is a critical temperature $T_{\it c}=\frac
{2J}{k\ln3}$. Below the critical temperature $T_c$, the
entanglement is increased when the magnetic field $B$ is
decreased. Above the critical temperature $T_c$, the entanglement
is always close to zero no matter the magnetic field $B$ is
increased or decreased.

\qquad For the entanglement teleportation of the whole two-qubit
system as the resource of the thermal mixed state in a 1D
Heisenberg $XXX$ chain, the standard teleportation through mixed
states can be regarded as a general depolarising channel [5, 6].
Similar to the standard teleportation, the entanglement
teleportation for the mixed channel of an input entangled state is
destroyed and its replica state appears at the remote place after
applying local measurement in the form of linear operators. When
two-qubit state $\rho_{in}$ is teleported via the channel, the
output state $\rho_{out}$ is [11]
\begin{equation}\label{5}
\rho_{out}=\sum_{ij}p_{ij}(\sigma_{i}\otimes\sigma_{j})\rho_{in}(\sigma_{i}\otimes\sigma_{j})
\ (\sum_{ij}p_{ij}=1).
\end{equation}
In the above equation, $\sigma_{i} (i=0, x, y, z)$ signify unit
matrix $I$ and are three components of Pauli matrix
$\overrightarrow{\sigma}$ correspondingly,
$p_{ij}=p_{i}p_{j}=tr(E^i\rho^c) \cdot tr(E^{j}\rho^{c})$,
$p_{00}$ stands for the maximal possibility of successful
teleportation, and $p_{i}$ =$tr(E^{i}$ $\rho^{c})$ is the ratio of
maximally entangled state to the thermal mixed states $\rho^{c}$
at the equilibrium temperature. Here
$\{E^{i}\}=\{|\psi^{-}\rangle\langle\psi^{-}|,|\phi^{-}\rangle\langle\phi^{-}|,|\phi^{+}\rangle\langle\phi^{+}|,
|\psi^{+}\rangle\langle\psi^{+}|\}$.

\qquad To see more clearly the entanglement teleportation of two
qubits, one kind of pure entangled state can be chosen as an
example
\begin{equation} \label{6}
|\psi\rangle_{in}=c_1|u_1\rangle|v_1\rangle+c_2|u_2\rangle|v_2\rangle,
\end{equation}
where $\{|u_1\rangle,|u_2\rangle\}$ and
$\{|v_1\rangle,|v_2\rangle\}$ are two sets of basis vectors of two
qubits. Without losing the generality, it can be assumed
\begin{equation}\label{7}
c_1=\cos\frac {\displaystyle \theta}{2} \quad\mbox{and}\quad
c_2=\sin\frac {\displaystyle \theta}{2}\exp(i\phi),
\qquad\mbox{when}\qquad \theta\in[0,\pi] \quad\mbox{and}\quad
\phi\in[0,2\pi].
\end{equation}
Here different values of $\theta$ describe all states with
different amplitudes, and $\phi$ stands for the phase of these
states. From Eqs. (3)-(7), the amount of entanglement of
$\rho_{out}$ can be expressed as,
\begin{equation}\label{8}
E_{out}=\max\{\frac {E_{in}(\exp(\frac
{2J}{T})-1)^{2}-4\cosh(\frac {B}{T})(\exp(\frac
{2J}{T})+1)}{[2\cosh(\frac {B}{T})+\exp(\frac
{2J}{T})+1]^{2}},0\},
\end{equation}
where $E_{out}$ is the entanglement of teleported state and
$E_{in}=2|c_1c_2|$ is the input entanglement.

\qquad  The quantity $E_{out}$ as a function of $E_{in}$ is
plotted in Fig. 1 when the magnetic field $B$, the temperature $T$
and the coupling coefficient $J$ are changed. Fig. 1(a) is a plot
of $E_{out}$ as functions of $E_{in}$ and $B$. When the input
entanglement $E_{in}=0$, $E_{out}$ is always zero no matter $B$ is
increased or not. The value of $E_{out}$ is decreased with
increasing the value of $B$. This is due to the fact that the
entanglement of $1D$ Heisenberg chain decreases when the magnetic
field is increased. It is seen that the entanglement of the
teleported state is closely related to that of the channel. While
$E_{out}$ is increased with increasing value of $E_{in}$. This is
due to their linear relationship in Eq. (8). Fig. 1(b) is a plot
of $E_{out}$ as functions of $E_{in}$ and $T$. The value of
$E_{out}$ is decreased with increasing value of $T$ since the
entanglement of the mixed channel decreases with $T$. While
$E_{out}$ is increased with increasing value of $E_{in}$. Fig.
1(c) is a plot of $E_{out}$ as functions of $E_{in}$ and $J$. It
is seen that $E_{out}$ is increased with increasing values of both
$J$ and $E_{in}$. This means that $E_{out}$ depends on the
coupling strength $J$. The higher value of $E_{out}$ is due to the
stronger coupling of $J$. It is found that $E_{out}$ is always
zero for $J\leq(\frac {\ln3}{2})T$ when there is no entanglement
in the mixed channel.

\qquad From Figs. 1(a) to 1(c), it is noted that $E_{out}$ is
increased linearly with increasing value of $E_{in}$. While
$E_{out}$ is decreased with increasing values of both $B$ and $T$.
Under the general circumstances, the output entanglement of
two-qubit state $|\psi\rangle_{in}$ will decrease via the quantum
channel. Since the entanglement cannot be increased under local
operations, Eq. (5) provides an upper bound to the output
entanglement. That is, the output entangled mixed state has the
entanglement smaller than that of the channel [5]. The
entanglement after the teleportation is always less than that of
the input state before the teleportation, i. e., $E_{out}<E_{in}$.
To realize the entanglement teleportation of $E_{out}>0$, the
quantum channel needs the minimal entanglement. Fig. 1(d) is a
plot of critical values of the magnetic field $B_c$, the
temperature $T_c$, and the coupling coefficient $J_c$ for
$E_{out}=0$ when $E_{in}=1$. The critical values of $B_c$, $T_c$,
and $J_c$ are located on the plane. When the values of $B$, $T$,
and $J$ are above the plane, the output entanglement is always
zero. That is, above these critical values, the entanglement
teleportation cannot be achieved. If the values of $B$, $T$, and
$J$ are located below the plane, $E_{out}>0$. That is, the
entanglement teleportation for the mixed channel can be realized.

\qquad The average fidelity $F_A$ of teleportation can be
formulated by
\begin{equation}\label{9}
F_A=\frac {\displaystyle \int_{0}^{2\pi}d\phi\!\int_{0}^{\pi}
F\,\sin\theta\,d\theta} {4\pi}.
\end{equation}
If a 1D Heisenberg $XXX$ chain is used as quantum channel, $F_A$
can be expressed as

\begin{equation}\label{10}
F_A=\frac 23\left\{1+\frac {\displaystyle \frac 52\exp(\frac
{4J}{T})+3\exp(\frac {2J}{T})+\frac 52-2\left[\exp(\frac
{2J}{T})+\cosh(\frac {B}{T})+1\right]^2} {\displaystyle
\left[\exp(\frac {2J}{T})+2\cosh(\frac {B}{T})+1\right]^2}
\right\}.
\end{equation}
In Eq. (10), the average fidelity $F_A$ is larger than $\frac 23$
when $T<\frac {2J}{k\ln11}$ and $B$ is not very large. This shows
that the teleportation through $1D$ Heisenberg mixed states is
better than the classical communication since $F_A=\frac 23$ is
the limited value in classical communication.

\qquad The average fidelity $F_A$ is plotted as functions of the
magnetic field $B$ and the coupling coefficient $J$ in Fig. 2 when
the temperature $T=\frac 1{2\ln3}$. From Fig. 2(a), it is seen
that $F_A$ is increased when $J$ is increased. When $J$ is small,
$F_A$ is gradually decreased and then increased with increasing
value of $B$. If the coupling is weak, $J\rightarrow0$, the
average fidelity can be approximated by
\begin{equation}
\label{11}
 F_{A}=(\frac 1{\cosh\frac BT+1}-\frac 13)^2+\frac 29.
\end{equation}
It is easily seen that there is a minimum fidelity of $F_A=\frac
29$ when $B=B_{m}=T\cosh^{-1}2$. The average fidelity $F_A$ can be
increased if the magnetic field $B\neq B_{m}$. Therefore, the
average fidelity $F_{A}$ in Fig. 2(a) is like a parabolic curve
when $B$ is increased from zero. If the interaction in the
Heisenberg chain is weak, the fidelity of teleportation may be
improved to some degree by increasing magnetic field. Fig. 2(b) is
a plot of the critical values of the magnetic field $B_{cf}$, the
temperature $T_{cf}$, and the coupling coefficient $J_{cf}$ for
$F_A=\frac 23$. From Fig. 2(b), it is seen that the critical
values of $B_{cf}$, $T_{cf}$, and $J_{cf}$ are located on the
plane. If the values of $B$, $T$, and $J$ are located on the plane
or below the plane, $F_A\geq\frac 23$. This means that the
entanglement teleportation of the mixed channel is superior to the
classical communication. If the values of $B$, $T$, and $J$ are
located above the plane, $F_A <\frac 23$. That is, above these
critical values, the fidelity $F_A$ will be less than $\frac 23$.
The entanglement teleportation will be worse.

\section{Mutual information of the quantum channel}

\qquad Compared with the classical information theory, the mutual
information $\mathcal{I}$ in the quantum communication theory can
imply the classical capacity of a quantum channel. For two-qubit
signal states, the value $2.0$ of mutual information means that
the classical information carried by signal states can be totally
transmitted via the quantum channel. If the value of $\mathcal{I}$
is $0.0$, it means that the original classical information coded
in signal states is totally destroyed after quantum teleportation.
If the value of $\mathcal{I}$ is $0.0<\mathcal{I}<2.0$, it means
that the original classical information coded in signal states is
destroyed partially after quantum teleportation. Given a set of
input signal states $\{q_i,\pi_i\}$, the output states are
$\{q_i,\chi_i\}$. Assuming $\chi=\sum_{i} q_i\chi_i$, the mutual
information can be written as [12-15],
\begin{equation}\label{12}
\mathcal{I}_{n}=S(\chi)-\sum_{i}q_{i}S(\chi_{i}) \
(\sum_{i}q_{i}=1),
\end{equation}
where the subscript $n$ denotes the number of the channel used. In
Eq. (12), the Von Neumann entropy is
$S(\rho)=-tr(\rho\log_2\rho)$. There are two independent 1D
Heisenberg chains as the quantum channel with $n=2$. The input
signal states can be written as
\begin{eqnarray}\label{13}
|\pi_1\rangle&=&\cos\gamma|00\rangle+\sin\gamma|11\rangle\nonumber,\\
|\pi_2\rangle&=&\sin\gamma|00\rangle-\cos\gamma|11\rangle\nonumber,\\
|\pi_3\rangle&=&\cos\beta|01\rangle+\sin\beta|10\rangle\nonumber,\\
|\pi_4\rangle&=&\sin\beta|01\rangle-\cos\beta|10\rangle.
\end{eqnarray}
Here, the possibility of each state is the same, i. e., $q_i=\frac
14$. The parameters $\gamma$ and $\beta$ describe two sets of
entangled states with different amplitudes. By means of Eqs. (5),
(12), and (13), the mutual information is

\begin{equation}\label{14}
\mathcal{I}=\mathcal{I}_2=2-\frac
14\sum_{ij}\xi_{ij}\log_2\xi_{ij},
\end{equation}
where $\xi_{ij}$ is the $j$th eigenvalue of $\chi_i$.

\qquad The mutual information $\mathcal{I}$ is illustrated in Fig.
3. Fig. 3(a) is a plot of the mutual information $\mathcal{I}$ as
functions of signal states $(\gamma, \beta)$. From Fig. 3(a), it
is seen that $\mathcal{I}$ is a periodic function of both $\gamma$
and $\beta$. There is a minimum mutual information of
$\mathcal{I}=1.70$ when the input signal states are maximally
entangled with $\ \beta=\gamma=(n+\frac 14)\pi$ or $ (n+\frac
34)\pi, (n=0, 1, 2, 3, ......)$. When the input states are not
entangled with $\ \beta=\gamma=n\pi$ or $(n+\frac 12)\pi, (n=0, 1,
2, 3, ......)$, the maximum mutual information of
$\mathcal{I}=1.80$ can be obtained. It is interesting to note that
the entanglement quality of the input states cannot enhance the
mutual information in entanglement teleportation. The difference
of mutual information $\mathcal{I}$ between maximally entangled
and non-entangled states is less than $6.0\%$. Fig. 3(b) is a plot
of the mutual information $\mathcal{I}$ as functions of the
magnetic field $B$ and temperature $T$ when the input signal
states are maximally entangled with $\ \beta=\gamma=(n+\frac
14)\pi$ or $ (n+\frac 34)\pi, (n=0, 1, 2, 3, ......)$. From Fig.
3(b), it is seen that high mutual information can be obtained when
input states are entangled with small values of magnetic field $B$
and temperature $T$.

\qquad The mutual information $\mathcal{I}$ is plotted in Fig. 4
when the input states $\beta$ and $\gamma$ are varied. From Fig.
4, it is seen that $\mathcal{I}$ monotonously decreases with
increasing values of both $B$ and $T$. Fig. 4(a) is a plot of
$\mathcal{I}$ as a function of $B$. In Fig. 4(a) of very strong
magnetic field $B$, the mutual information $\mathcal{I}$ of
non-entangled input states (solid line) is decreased more rapidly
than that of maximally entangled ones (dashed line). That is, the
mutual information of non-entangled input states is more sensitive
to the magnetic field $B$ than that of maximally entangled ones.
When the magnetic field is stronger than $1.3$, the mutual
information of maximally entangled states decreases slower than
that of non-entangled states. Fig. 4(b) is a plot of $\mathcal{I}$
as a function of $T$. From Fig. 4(b), it is seen that there is
almost no difference in $\mathcal{I}$ between maximally entangled
input state and non-entangled input state when $T<0.2$. When
$T>0.2$, the mutual information of non-entangled input states
(solid line) is always higher than that of maximally entangled
ones (dashed line).

\section{Discussion}

\qquad The entanglement teleportation of two-qubit mixed states
via two independent 1D Heisenberg $XXX$ chains is analyzed. For
suitable values of the magnetic field $B$, the temperature $T$,
and the coupling coefficient $J$, the minimal entanglement of the
thermal state in 1D Heisenberg chain is needed to realize the
entanglement teleportation. There exist critical values of $B$,
$T$, and $J$ when the minimal entanglement is obtained. Above the
critical values, the entanglement teleportation cannot be
achieved. When $T<\frac {2J}{\ln11}$ and $B$ is small, the average
fidelity is larger than $2\!/3$ which is better than the classical
communication. To keep high value of the mutual information, the
magnetic field and the temperature should be very small. The
quality of the entanglement of input states cannot enhance the
mutual information. For maximally entangled input states, minimal
value of mutual information is obtained. While for non-entangled
input states, maximal mutual information is achieved.

\vskip 0.4cm

{\large \bf Acknowledgement}

\qquad It is a pleasure to thank Professor Yinsheng Ling for his
many helpful discussions.

\newpage

{\Large \bf Reference}

1. C. H. Bennett, G. Brassard, C. Cr$\acute{e}$peau, R. Jozsa, A.
Peres, and W. K.

\ \ \ \ Wootters, Phys. Rev. Lett. {\bf 70}, 1895(1993).

2. A. K. Ekert, Phys. Rev. Lett. {\bf 67}, 66(1991).

3. M. C. Arnesen, S. Bose and V. Vedral, Phys. Rev. Lett. {\bf
87}, 017901(2001).

4. X. Wang, Phys. Rev. {\bf A64}, 012313(2001).

5. G. Bowen and S. Bose, Phys. Rev. Lett. {\bf 87}, 267901(2001).

6. M. Horodecki, P. Horodecki and R. Horodecki, Phys. Rev. {\bf
A60}, 1888(1999).

7. D. Bru$\ss$, L. Faoro, C. Macchiavello and G. M. Palma, J. Mod.
Opt. {\bf 47},

\ \ \ \ 325(2000).

8. J. Lee and M. S. Kim, Phys. Rev. Lett. {\bf 84}, 4236(2000).

9. Y. Yeo, Phys. Rev. {\bf A66}, 062312(2002).

10. M. Horodecki, P. Horodecki and R. Horodecki, Phys. Lett. {\bf
A223}, 1(1996).

11. A. Peres, Phys. Rev. Lett. {\bf 70}, 1413(1996).

12. C. Macchiavello, G. M. Palma and S. Virmani, quant-ph/0307016.

13. H. Barnum, M. A. Nielsen and B. Schumacher, Phys. Rev. {\bf
A57}, 4153(1998).

14. C. Macchiavello and G. M. Palma, Phys. Rev. {\bf A65},
050301(2002).

15. B. Schumacher, Phys. Rev. {\bf A54}, 2614(1996).

\newpage

{\Large \bf Figure Captions}

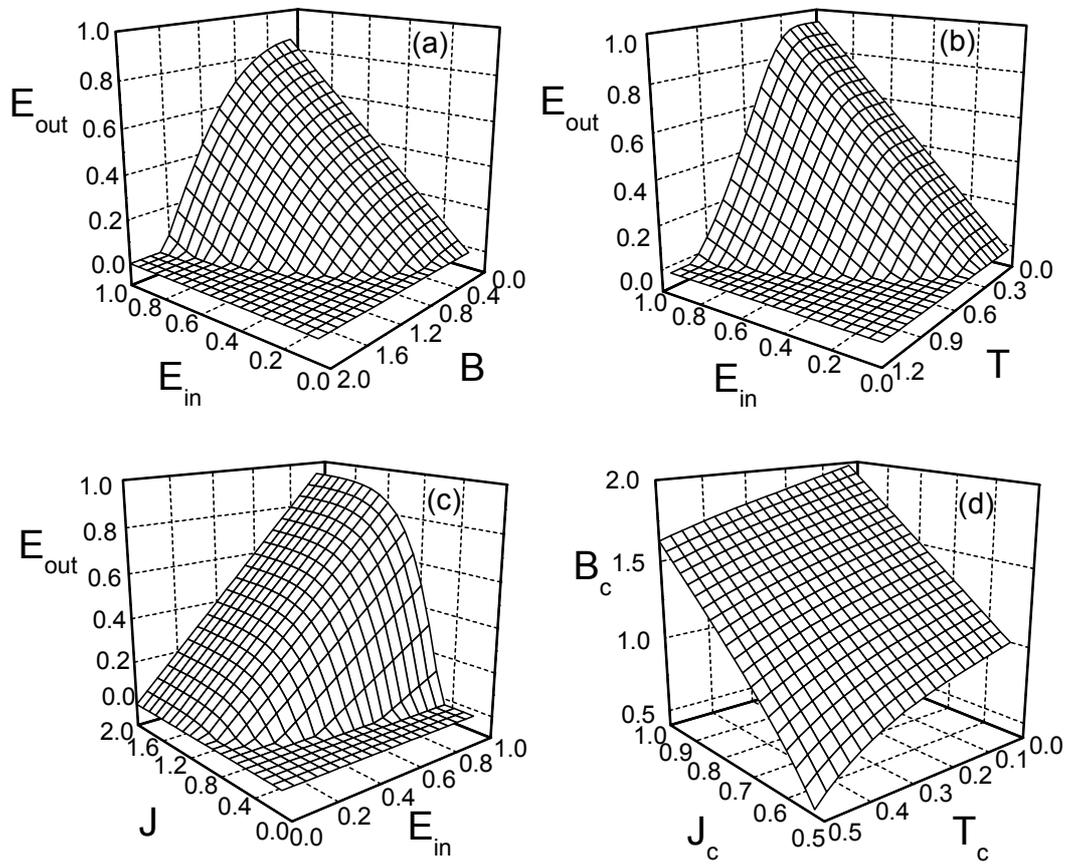
{\bf Fig. 1.}

The teleported entanglement $E_{out}$ as a function of the input
entanglement $E_{in}$ if

(a). the magnetic field $B$ is changed when $J=1$ and $T=\frac
1{2\ln3}$;

(b). the temperature $T$ is changed when $J=1$ and
$B\rightarrow0$;

(c). the coupling coefficient $J$ is changed when $T=\frac
1{2\ln3}$ and $B\rightarrow0$.

(d). The critical values of $B_c$, $T_c$, and $J_c$ for
$E_{out}=0$ when $E_{in}=1$.

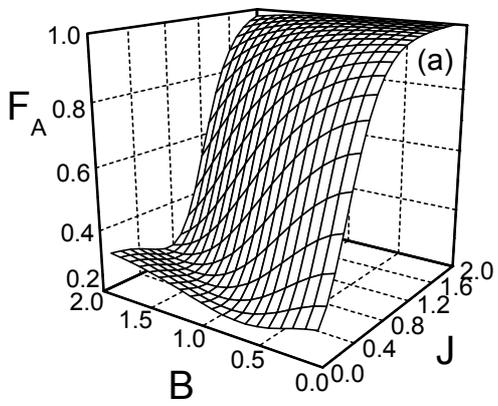
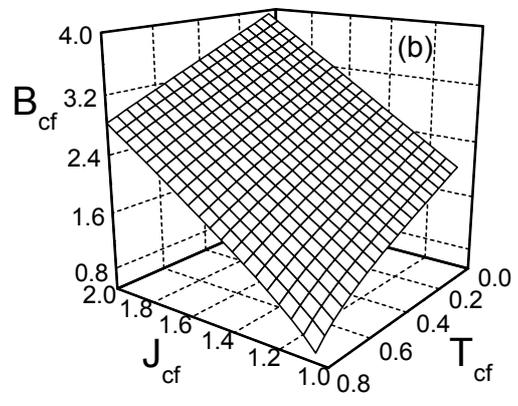
{\bf Fig. 2.}

(a). The average fidelity $F_A$ as functions of the magnetic field
$B$ and the coupling coefficient $J$ when the temperature $T=\frac
1{2\ln3}$.

(b). The critical values of $B_{cf}$, $T_{cf}$, and $J_{cf}$ when
$F_A=\frac 23$.

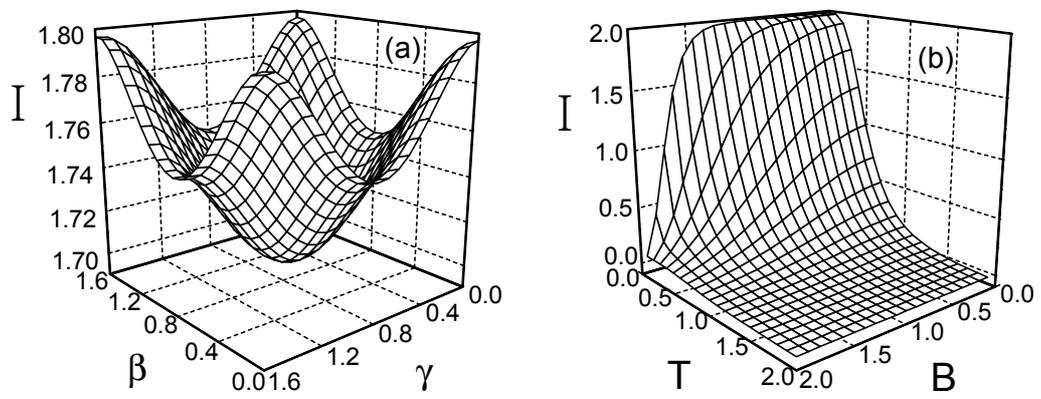
{\bf Fig. 3.}

The mutual information $\mathcal{I}$ as functions of the input
signal entangled states $(\beta,\gamma)$ and parameters $(B, T)$.

(a). $\mathcal{I}$ as functions of $(\beta,\gamma)$ when $J=1$,
$T=\frac 1{2\ln3}$, and $B\rightarrow0$.

(b). $\mathcal{I}$ as functions of $(B, T)$ when $J=1$ and
$\beta=\gamma=(n+\frac 14)\pi$ or $ (n+\frac 34)\pi, (n=0, 1, 2,
3, ......)$.

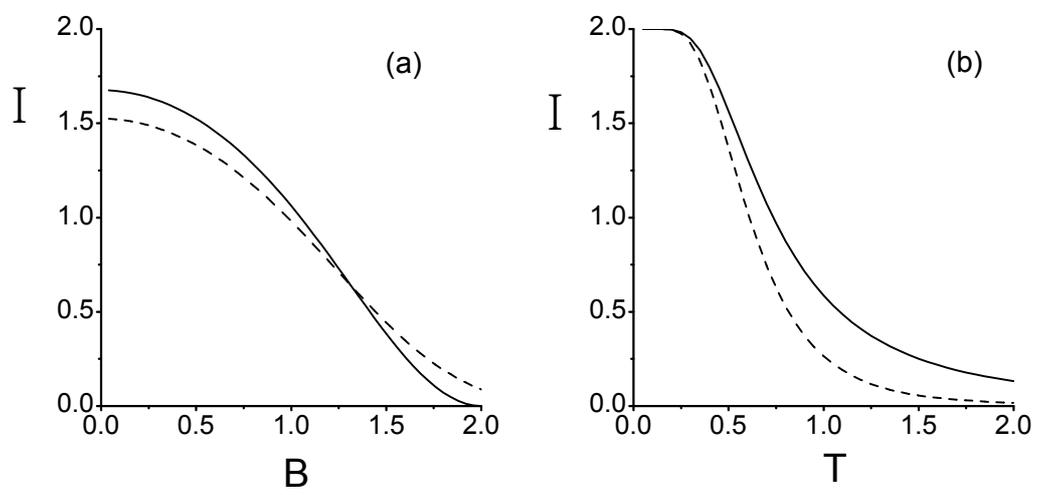
{\bf Fig. 4.}

The mutual information $\mathcal{I}$ as a function of the magnetic
field $B$ and the temperature $T$.

(a). $\mathcal{I}$ as a function of $B$ when $J=1$ and $T=\frac
1{2\ln3}$.

(b). $\mathcal{I}$ as a function of $T$ when $J=1$ and
$B\rightarrow0$.

The dash line labels maximally entangled input states when
$\beta=\gamma=(n+\frac 14)\pi$ or $ (n+\frac 34)\pi, (n=0, 1, 2,
3, ......)$. The solid line labels non-entangled input states when
$\beta=\gamma=n\pi$ or $(n+\frac 12)\pi, (n=0, 1, 2, 3, ......)$.

\end{document}